# Graph-based vulnerability assessment of resting-state functional brain networks in full-term neonates

Mahshid Fouladivanda[1], Kamran Kazemi[1], Habibollah Danyali[1], Ardalan Aarabi[2,3*]

[1] Department of Electrical and Electronics Engineering, Shiraz University of Technology, Shiraz, Iran
[2] Laboratory of Functional Neuroscience and Pathologies (UPJV UR 4559), University Research Center (CURS), University of Picardy Jules Verne, Amiens, France
[3] Faculty of Medicine, University of Picardy Jules Verne, Amiens, France

* Correspondence to arladan.aarabi@u-picardie.fr

**Abstract**

Network disruption during early brain development can result in long-term cognitive impairments. In this study, we investigated rich-club organization in resting-state functional brain networks in full-term neonates using a multiscale connectivity analysis. We further identified the most influential nodes, also called spreaders having higher impacts on the flow of information throughout the network. The network vulnerability to damage to rich-club (RC) connectivity within and between resting-state networks was also assessed using a graph-based vulnerability analysis.

Our results revealed a rich club organization and small-world topology for resting-state functional brain networks in full-term neonates regardless of the network size. Interconnected mostly through short-range connections, functional rich-club hubs were confined to sensory-motor, cognitive-attention-salience (CAS), default mode, and language-auditory networks with an average cross-scale overlap of 36%, 20%, 15% and 12%, respectively. The majority of the functional hubs also showed high spreading potential, except for several non-RC spreaders within CAS and temporal networks. The functional networks exhibited high vulnerability to loss of RC nodes within sensori-motor cortices, resulting in a significant increase and decrease in network segregation and integration, respectively. The network vulnerability to damage to RC nodes within the language-auditory, cognitive-attention-salience, and default mode networks was also significant but relatively less prominent. Our findings suggest that the network integration in neonates can be highly compromised by damage to RC connectivity due to brain immaturity.

**Introduction**

Advances in network analysis have provided new insights into the structural and functional organization underpinning the human brain connectome (Sporns et al. 2005; Bullmore and Sporns 2009). The focus of network studies has been on quantifying measures of network segregation and integration of information. In the past decade, there has been growing interest in identifying "cortical hubs" shown to play a critical role in information integration at the network level. In adults, the existence of densely connected hubs (so-called "rich club") has been widely investigated in structural and functional brain networks (Chklovskii et al. 2002; Kaiser and Hilgetag 2006; Hagmann et al. 2008; Van Den Heuvel and Pol 2010; Bullmore and Sporns 2012; van den Heuvel and Sporns 2013; Fukushima et al. 2018). Previous studies have consistently shown that damage to RC hubs can cause widespread disruption across large-scale brain networks with a significant impact on cognition and behavior due to the central role of the rich-club in global communication between brain regions (Van Den Heuvel et al. 2012; Collin et al. 2014).

Compared to the adult brain, our knowledge of the brain's structural and functional organization in the early stages of neurodevelopment is still limited (Amsterdam 1972; Haith et al. 1988; Reznick 2009). Many studies have suggested that network disruption during brain development can result in long-term cognitive deficits (Grayson et al. 2014; Gao et al. 2015; Van Den Heuvel et al. 2015; Cao et al. 2017a, b; Gozdas et al. 2019). Several network studies have reported findings concerning the maturity of sensory-motor networks at birth, suggested to be related to the basic survival functions highly contributing to the gradual maturation of higher-level integrative areas at later stages of brain development (Guillery 2005; Hagmann et al. 2010; Yap et al. 2011; Bullmore and Sporns 2012; Buckner and Krienen 2013; Tymofiyeva et al. 2013; Collin et al. 2014; Vértes and Bullmore 2015; Cao et al. 2017a; Bruchhage et al. 2020). In full-term neonates, a few network studies have reported findings concerning rich-club hubs predominately found along the anterior-posterior medial axis of the cortex (Ball et al. 2014; Van Den Heuvel et al. 2015; Fouladivanda et al. 2021). Since rich-club hubs are highly interconnected, their integrative role in cross-linking different brain regions and networks makes them indispensable for maintaining efficient brain functioning. It is suggested that the brain immaturity at birth can significantly alter functional rich club organization by reducing connections between RC nodes, resulting in cognitive deficits in later stages of neurodevelopment (Scheinost et al. 2016b). However, the extent to which brain immaturity can affect functional connectivity in neonates is still poorly understood.

In the present study, we assessed the whole-brain functional connectome and the resilience of neonatal functional brain networks to damage to resting-state rich-club connectivity using a graph-based vulnerability analysis. To this end, the network topology was first investigated at five coarse to fine parcellation scales. We then identified rich-club nodes and connections as well as the most influential nodes (spreaders) that participated in resting-state functional networks involved in primary and high-level cognitive processes. Finally, we assessed network vulnerability under rich club node and connection attacks using graph measures of integration and segregation.

## Materials and Methods

**Subjects and MR Acquisition**

MR images of hundred full-term neonates (37-44 weeks gestational age at scan) from the Developing Human Connectome Project dataset (dHCP, second release, developingconnectome.org) were included in this study. In this dataset, multimodal MR images have been collected from all neonates during natural sleep after gathering informed parental consent using a 3T Philips Achieva imaging system and a dedicated 32-channel neonatal phased-array head coil (Hughes et al. 2017). The dHCP study was approved by the UK Health Research Authority. The second release of dHCP includes anatomical MR images containing high-resolution 3D T1w and T2w images with an in-plane resolution of $0.8 \times 0.8$ mm$^2$ and a thickness of 1.6 mm overlapped by 0.8 mm. Resting-state fMRI (rsfMRI) data were also acquired using a high temporal resolution multiband EPI sequence (TR=392ms; 2.15mm isotropic) adapted for neonates (Price et al. 2015). Single-band EPI reference scans were collected with bandwidth-matched readout followed by spin-echo EPI acquisitions with (Posterior ⇔ Anterior) phase-encoding directions. More details about the data acquisition can be found in (Fitzgibbon et al. 2020).

**Data preprocessing**

The T2w images were analyzed by the dHCP minimal preprocessing procedure including bias correction, brain extraction, and segmentation into gray matter (GM), white matter (WM), and Cerebrospinal fluid (CSF) (Makropoulos et al. 2018). The rsfMRI data were also analyzed by the dHCP minimal preprocessing procedure (Fitzgibbon et al. 2020) including corrections and denoising for susceptibility field distortion, slice timing, and slice-to-volume motion artifacts (Salimi-Khorshidi et al. 2014). The functional images were then registered to the standard 40-week template space (Serag et al. 2012) using the T2w images. In addition, we regressed out the averaged WM and CSF signals from each GM voxel's time series for each neonate. The rsfMRI data were finally bandpass filtered (0.009-0.08 Hz) and spatially smoothed using a Gaussian kernel (FWHM = 3mm) (Howell et al. 2020).

**Resting-State network analysis**

The preprocessed rsfMRI data were decomposed into 60 independent components (IC) using a group ICA as implemented in Group ICA of fMRI Toolbox (GIFT) (Calhoun et al. 2001; Rachakonda et al. 2007; Erhardt et al. 2011). After removing artifact/noise ICs exhibiting peak activations in white matter or high spatial overlap with vascular, ventricular, motion, and susceptibility artifacts, the remaining components were grouped into eight RSNs based on prior knowledge reported in other studies (Smith et al. 2009; Thomas Yeo et al. 2011; Fouladivanda et al. 2021). To quantify the overlap between each parcel and RSNs (functional modules), each parcel was assigned to each functional module, if its overlap exceeded 3% based on the "winner-take-all" strategy (van den Heuvel and Sporns 2013).

**Functional network construction**

We used the spatially-constrained normalized-cut (NCUT) spectral clustering algorithm (Craddock et al. 2012) to perform functional parcellation at five spatial scales comprising 100, 300, 500, 700, and 900 parcels within cortical and subcortical regions, and cerebellum (Fouladivanda et al. 2021). For each neonate, we constructed five functional connectivity matrices, in which the edge weights were computed using Pearson's correlation between the average time courses of voxels within the parcels. A group average correlation matrix was then computed at each scale by keeping both positive and negative correlations as suggested in other studies (Goelman et al. 2014; Thompson and Fransson 2015). To construct undirected networks, the group matrices were first thresholded over a range of densities within [0.05 0.4] with an increment of 0.05 (Bassett and Bullmore 2006). Across all scales, an optimal proportional threshold was then determined by maximizing the global cost-efficiency of the brain networks (Bassett et al. 2009) to reduce the number of false-positive edges and minimize the effect of noise (Drakesmith et al. 2015). The group correlation matrices were finally thresholded using the optimal threshold (herein 0.2) to obtain adjacency connectivity matrices.

**Network topological properties**

To examine the dependency of the whole-brain functional network topology on the nodal scale, we computed commonly-used global and local graph metrics including degree, clustering coefficient, path length, global and local efficiency, betweenness centrality, assortativity, modularity, and small-worldness (Freeman et al. 1979; Watts and Strogatz 1998; Amaral et al. 2000; Latora and Marchiori 2001; Newman 2002, 2006). To control for network size (Sporns 2010), the normalized clustering coefficient and the normalized path length were also computed for each network with reference to 10,000 random graphs.

**Rich club analysis**

For each scale, the rich-club organization of the group-average functional brain network was investigated over a range of k levels using the normalized rich-club coefficient ($\varphi_{norm}$) computed with reference to the average rich-club coefficient of 10,000 random networks of equal size with similar connectivity distributions (van den Heuvel and Sporns 2011; Khalilian et al. 2021). To compare the rich club organization across different scales, the k level with $\varphi_{norm} > 1$ was selected for each scale such that the top 20% of the group network nodes with the highest degrees were in the rich club. To identify cross-scale rich-club regions in the automated anatomical atlas (AAL) (Tzourio-Mazoyer et al. 2002), a rich club score was first computed for each AAL region based on the proportion of the overlap between the region and the rich club nodes. The average rich-club score was then computed over the scales to identify cross-scale high-ranked rich-club regions (Fouladivanda et al. 2021).

Group differences between RC/non-RC nodes and connections (RC vs feeder/local connections) were assessed using permutation testing with 10,000 permutations (Fouladivanda et al. 2021). For each permutation, the Erdos-Reiniy random matrices were generated with randomized class assignments to obtain an empirical null distribution of effects under the null hypothesis with p-values computed as the percentage of the computed values exceeding the empirically obtained values (van den Heuvel and Sporns 2013). A p-value less than 0.05 was considered statistically significant.

**Most Influential Nodes**

Among highly-connected hubs, it is shown that the so-called "spreaders" located in the core of the network have higher impacts on the flow of information throughout the network (Kitsak et al. 2010). The spreading potential of highly connected hubs is associated with their betweenness centrality. Therefore, highly connected hubs do not necessarily correspond to the best spreaders across the network. To identify the most influential nodes with high spreading potential, Salavaty et al. (2020) introduced an algorithm termed Integrated Value of Influence (IVI) by combining the most important local and global network centrality measures. In this method, the IVI score of each node *i* computed based on the minimum cost-maximum efficiency criteria (Salavaty et al. 2020) as:

$$\text{IVI}_i = (\text{Hubness}_i) \times (\text{Spreading}_i) \quad (1)$$

$$\text{Hubness}_i = D_i + H_i$$

$$\text{Spreading}_i = (NC_i + CR_i) \times (BC_i + Cl_i)$$

Where, as local centrality measures, $D_i$ is the nodal degree, $H_i$ is a semi-local centrality index (Lü et al. 2016) indicating the existence of minimum *h* neighbors for node *i* with a degree greater or equal to *h*, and $NC_i$ is the average neighborhood connectivity of node *i* (Maslov and Sneppen 2002). The Cluster Rank ($CR_i$) (Chen et al. 2013) is described based on the local centrality effect of the node ($f(c_i)$), and nodes in the neighborhood of node i ($\Gamma_i$) as $CR_i = f(c_i) \sum_{j \epsilon \Gamma_i}(D_j^{out} + 1)$. The $BC_i$ and $Cl_i$ are the global centrality measures determined based on the Betweenness centrality and collective influence, respectively (Freeman et al. 1991; Morone and Makse 2015). The nodal tendency to have the shortest path to other nodes of the network describes $BC_i$ (Salavaty et al. 2020). The collective information flow influence of node *i* in the entire network is defined by $Cl_i = (D_i - 1)\sum_{j \epsilon \delta B(i,l)}(D_j - 1)$ where $\delta B(i,l)$ are nodes with *l* distance from node i (Salavaty et al. 2020). All the centrality measures used in *IVI* are range normalized. In our study, we first computed the IVI for each node at different scales. We then selected the top 20% of the nodes with highest IVI to investigate their spatial overlap with the RC spatial maps.

**Vulnerability analysis**

To assess the vulnerability of the functional networks to damage to rich club connectivity in neonates, we conducted a series of simulations using random (1,000 randomizations) and targeted attacks on RC nodes and connections. The vulnerability index (*V*) was defined based on alterations in network measures (p) (Costa et al. 2007) as :

$$V = \frac{P_N - P_i}{P_N} \times 100 \qquad (2)$$

where $P_N$ and $P_i$ were the network degree, betweenness, global efficiency, clustering coefficient, and modularity before and after the attack, respectively.

**Attack on rich club nodes**: The network resilience to full attacks on RC nodes within each RSN was investigated by removing them and assessing changes in the network vulnerability index at each spatial scale.

**Attack on connections of RC nodes:** The network vulnerability to damage to connections of RC nodes was evaluated with random and targeted attacks with respect to the rich club (k level) range with φ$_{norm}$ > 1 for each scale. We first hypothesize that brain immaturity in neonates can weaken the degree of rich club nodes while maintaining their rich-clubness within the rich club range. To simulate this condition, the connections of rich club nodes were randomly removed to reduce their degree to the lowest k level within the rich club range at full (100%) targeted attacks on RC nodes within each RSN. Based on the hypothesis that brain immaturity can reduce the degree of rich club nodes below the lowest rich club (k) level in each scale, the connections of all RC nodes in each RSN were attacked to reduce their degree below the minimum rich club level. The average vulnerability index was then computed for the whole brain network for each condition.

To further investigate the extent of alterations in network measures due to random attacks on local nodes, the degree of N randomly selected local nodes was reduced by removing their local connections, where N was a random number between the minimum and the maximum number of RC nodes across all RSNs. In this simulation, RC and feeder connections were retained to keep the functional rich club intact across the entire network. The average vulnerability index was then computed over 1,000 random attacks.

**Results**

**Brain network properties**

Most of the global and local metrics remained relatively stable with a network size of at least 300 nodes (Table 1). The assortativity decreased significantly at finer nodal scales. An increasing trend was found for the mean nodal degree and betweenness.

**Table 1.** Global and local graph measures (median ± standard error) of the functional brain networks across all neonates.

| | | Nodal scale | | | | |
|---|---|---|---|---|---|---|
| | | **100** | **300** | **500** | **700** | **900** |
| **Global metrics** | **Small-worldness** | 1.67 ± 0.03 | 1.84 ± 0.02 | 1.85 ± 0.02 | 1.84 ± 0.02 | 1.83 ± 0.02 |
| | **Normalized Clustering Coefficient** | 2.69 ± 0.024 | 2.65 ± 0.027 | 2.60 ± 0.029 | 2.59 ± 0.029 | 2.59 ± 0.03 |
| | **Clustering Coefficient** | 0.53 ± 0.005 | 0.53 ± 0.005 | 0.52 ± 0.006 | 0.52 ± 0.006 | 0.52 ± 0.006 |
| | **Transitivity** | 0.55 ± 0.007 | 0.52 ± 0.007 | 0.51 ± 0.007 | 0.50 ± 0.007 | 0.50 ± 0.007 |
| | **Normalized Path Length** | 1.54 ± 0.06 | 1.40 ± 0.03 | 1.41 ± 0.02 | 1.41 ± 0.02 | 1.42 ± 0.01 |
| | **Global Efficiency** | 0.54 ± 0.005 | 0.57 ± 0.003 | 0.58 ± 0.002 | 0.59 ± 0.001 | 0.59 ± 0.001 |
| | **Modularity** | 0.26 ± 0.004 | 0.22 ± 0.003 | 0.20 ± 0.003 | 0.19 ± 0.003 | 0.19 ± 0.003 |
| | **Assortativity** | 0.16 ± 0.01 | 0.13 ± 0.01 | 0.085 ± 0.01 | 0.07 ± 0.01 | 0.06 ± 0.01 |
| **Local metrics** | **Degree** | 19 ± 0.13 | 54 ± 0.22 | 103 ± 0.31 | 134 ± 0.35 | 163 ± 0.38 |
| | **Local Efficiency** | 0.76 ± 0.002 | 0.75 ± 0.001 | 0.75 ± 0.001 | 0.75 ± 0.0004 | 0.75 ± 0.0003 |
| | **Betweenness centrality** | 62.65 ± 1.06 | 188.19 ± 1.68 | 340.77 ± 2.23 | 423.77 ± 2.51 | 506.75 ± 2.75 |

**Rich club organization**

The functional architecture of the neonatal brain displayed a rich club organization for k levels (with $\varphi_{norm}(k) > 1$, p< 0.05, 10,000 permutations) within [7-42], [15-155], [9-329], [15-439] and [16-539] for the nodal scale of 100, 300, 500, 700 and 900, respectively. As shown in Fig. 1, the majority of the top 20% of nodes exhibiting both high degree and high IVI was found in the bilateral central, temporal, and parietal regions as well as along the anterior-posterior medial axis of the cortex at different nodal scales. The nodes characterized by either high IVI or high rich-clubness were mostly located in the frontal and temporal lobes.

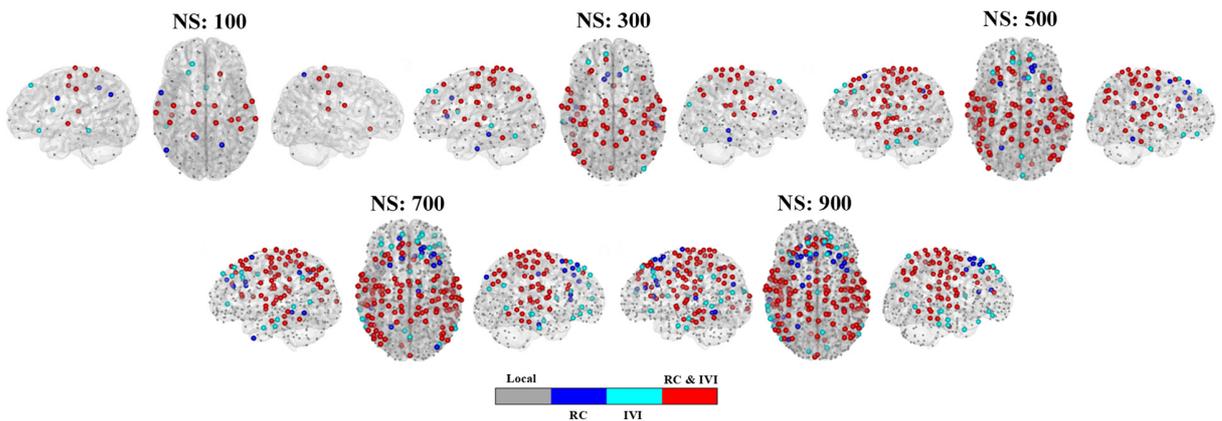

**Figure 1.** Spatial maps of the top 20% of nodes with high degree (rich-clubness, blue), high IVI (cyan) or both (Red) at each nodal scale (NS). Nodes represent the center-of-mass of functional parcels.

Fig. 3 illustrates the spatial overlap between the top 20% of RC/IVI nodes and RSNs (Fig. 2, Table 2) including Sensory-Motor Network (SM), Language and Auditory Network (LAN), Default-Mode Network (DMN), Visual Network (VIS), Cognitive-Attention-Salience Network (CAS), Temporal Network (TMN, middle and inferior temporal regions), Cerebellum (CB) and Sub-Cortical network (SC). The SM and CAS highly overlapped with the rich-club and nodes with high IVI with an average overlap of 35.56% (± 10.79%), and 20.04% (± 4.94%) across all scales, respectively. The DMN (14.97 ± 1.5%), LAN (11.5 ± 2.4%), SC (8.75 ± 0.5%), and TMN (7.52 ± 4.3%) also showed involvement in the functional rich club with lower contributions. The visual network exhibited very low spatial overlap with the rich club with a cross-scale mean of 1.65% (± 1.03%). The cerebellum also showed no or little involvement in the functional rich club at all scales. The main differences between the RC and IVI spatial maps were observed within CAS and TMN.

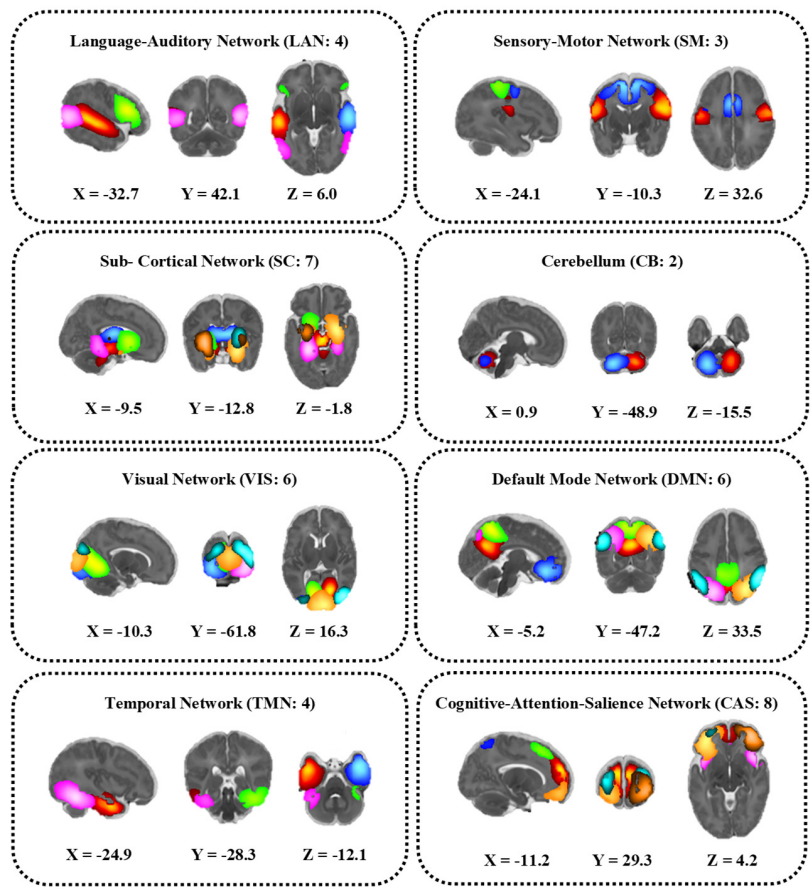

**Figure 2.** Spatial t-statistics maps of independent components grouped into eight resting-state networks (RSNs). For each IC, t values range from $t_{min}=3$ (darker colors) to $t_{max}$ (listed in Table S1, lightest colors).

**Table 2.** Peak activation of independent components (ICs) grouped in eight functional modules.

| Resting state network | IC, AAL region | $t_{max}$ | MNI coordinate (mm) |
|---|---|---|---|
| **Language-Auditory Network (LAN)** | IC2, Temporal_Sup_L | 25.24 | -30, -28, 14 |
| | IC7, Temporal_Sup_R | 25.8 | 34, -24, 14 |
| | IC19, Temporal_Sup_L | 15.8 | -30, 4, 18 |
| | IC38, Temporal_Mid_R | 14.95 | 26, -48, 14 |
| **Sensory-Motor network (SM)** | IC5, Postcentral_R | 14.9 | 50, -64, 20 |
| | IC16, Supp_Motor_Area_L | 14.3 | -4, --6, 44 |
| | IC37, Supp_Motor_Area_R | 12.5 | 16, -24, 50 |
| **Cognitive-Attention-Salience network (CAS)** | IC18, Cingulum-Ant_L | 18.35 | -6, 30, 18 |
| | IC24, Parietal_Inf_L | 15.2 | -26, -30, 40 |
| | IC28, Frontal_Mid_R | 15.34 | 18, 8, 38 |
| | IC29, Cingulum_Mid_R | 12.2 | 2, 12, 22 |
| | IC33, Frontal_Inf_Tri_L | 18.28 | -20, 42, 2 |
| | IC35, Frontal_Mid_R | 20.6 | 20, 26, 22 |
| | IC36, Frontal_Inf_Orb_R | 20.7 | 22, 26, 2 |
| | IC39, Parietal_Inf_R | 16.4 | 24, -30, 40 |
| **Temporal Network (TMN)** | IC8, Temporal_Inf_L | 23.2 | -32, -12, -8 |
| | IC10, Temporal_Mid_R | 24.5 | 24, -4, -8 |
| | IC21, Temporal_Inf_R | 17 | 32, -40, 2 |
| | IC22, Temporal_Mid_L | 19.5 | -34, -36, 2 |
| **Sub-Cortical network (SC)** | IC12, BrainStem | 18.3 | 0, -70. -2 |
| | IC14, Thalamus_R | 19.7 | 4, -16, 18 |
| | IC20, Putamen_L | 21.6 | -12, 0, 8 |
| | IC25, Hippocampus_L | 18.7 | -14, -30, 0 |
| | IC27, Hippocampus_R | 20 | 14, -8, -2 |
| | IC30, Putamen_R | 17.5 | 18, 0, 14 |
| | IC32, Putamen_L | 15.3 | -16, -18, 10 |
| **Default Mode Network (DMN)** | IC4, Precuneus_R | 19.4 | 2, -42, 28 |
| | IC13, Cingulum_Ant_L | 18.72 | -2, 24, -4 |
| | IC17, Cingulum_Post_L | 17.8 | -4, -38, 40 |
| | IC26, Precuneus_L | 18.1 | -16, -56, 26 |
| | IC31, Angular_R | 18.3 | 18, -50, 34 |
| | IC40, Angular_L | 12.2 | -32, -40, 30 |
| **Visual network (VIS)** | IC6, Lingual_R | 23.4 | 12, -48, 12 |
| | IC9, Occipital_Mid_L | 25.1 | -16, -66, 6 |
| | IC11, Lingual_L | 23.4 | -8, -50, 10 |
| | IC15, Occipital_Inf_R | 25.4 | 20, -64, 8 |
| | IC23, Calcarine_R | 20.5 | 2, -68, 14 |
| | IC34, Occipital_Mid_R | 13.6 | 26, -58, 16 |
| **Cerebellum (CB)** | IC1, Cerebellum | 28.9 | -12, -44, -10 |
| | IC3, Cerebellum | 26.7 | 10, -46, -12 |

Anatomical Automatic Labeling; $t_{max}$ = maximum t-statistic in each cluster; Coordinate = coordinate (mm) of peak activation in MNI space.

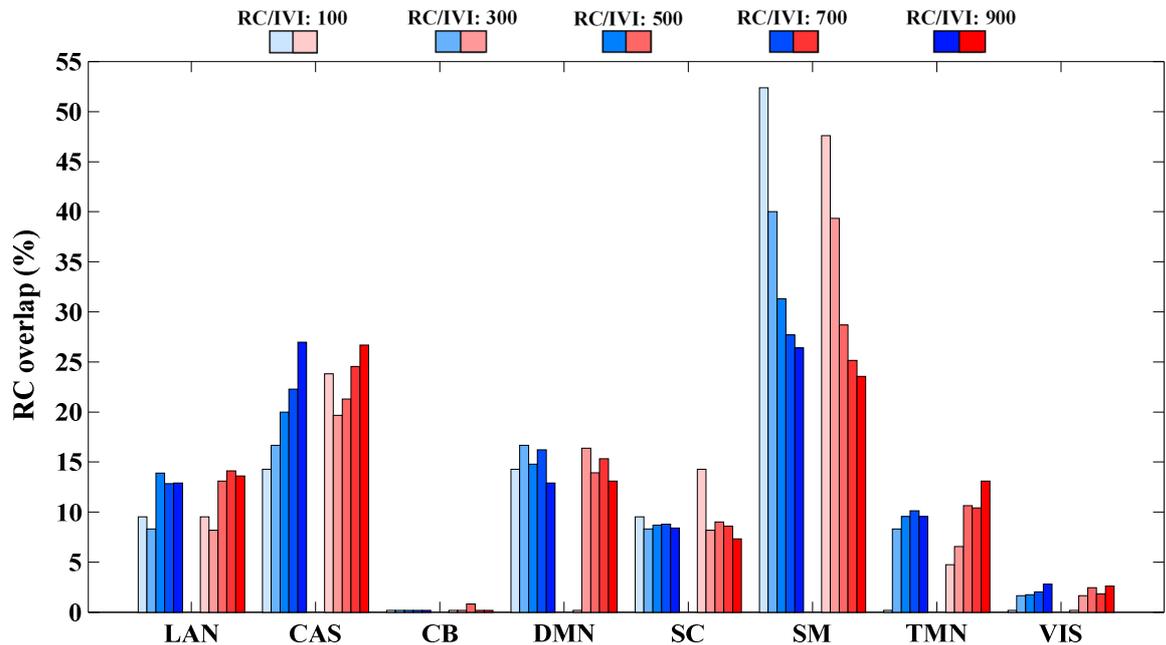

**Figure 3.** Overlap between the top 20% of RC/IVI nodes and RSNs at each nodal scale.

On average, the majority of the nodes within SM (82.4 ± 7.4%) were in the rich club across different scales. Almost one-fourth of the nodes in LAN (25.1 ± 5.6%) and DMN (23.7 ± 2.96%) showed rich-clubness. The cross-scale rich club proportions were lower for CAS (19.2 ± 5.1%), TMN (14.5 ± 8.2%), SC (8.3 ± 1.3%) and VIS (4.9 ± 3.4%).

Only 26% (± 15.1%) of the inter-RSN connections were long-range rich club or feeders across all scales. The intra-RSN connections were mostly short-range and local (54.2 ± 1.7%). Interestingly, the mean physical length of feeders was 23.2 mm (±8.9 mm), longer than that of the rich club (21.3 ± 8.5 mm) and local connections (13.85± 8.8 mm). The proportions of the short and long-range rich club, feeder, and local connections remained relatively stable across different scales.

**Rich club and influential regions**

Fig. 4 illustrates the high-ranked rich club regions across all scales. As shown, precentral and postcentral gyri (PreCG and PoCG), superior, middle and inferior temporal gyri (STG, MTG, and ITG), median cingulate (DCG), precuneus (PCUN), angular gyri (ANG), supramarginal gyrus (SMG), supplementary motor area (SMA), rolandic operculum (ROL), inferior parietal lobule (IPL) and insula (INS) displayed high rich club scores across all scales. Some regions such as the middle frontal gyrus (MFG), anterior cingulate gyrus (ACG) and caudate (CAU) showed higher RC scores at finer nodal scales. The paracentral lobule (PCL), superior parietal gyrus (SPG), and middle occipital gyrus (MOG) were, however, identified as high-degree RC nodes only at the nodal scales comprising less than 500 nodes. Across all scales, the RC regions also exhibited high IVI, however, their priority changed based on IVI.

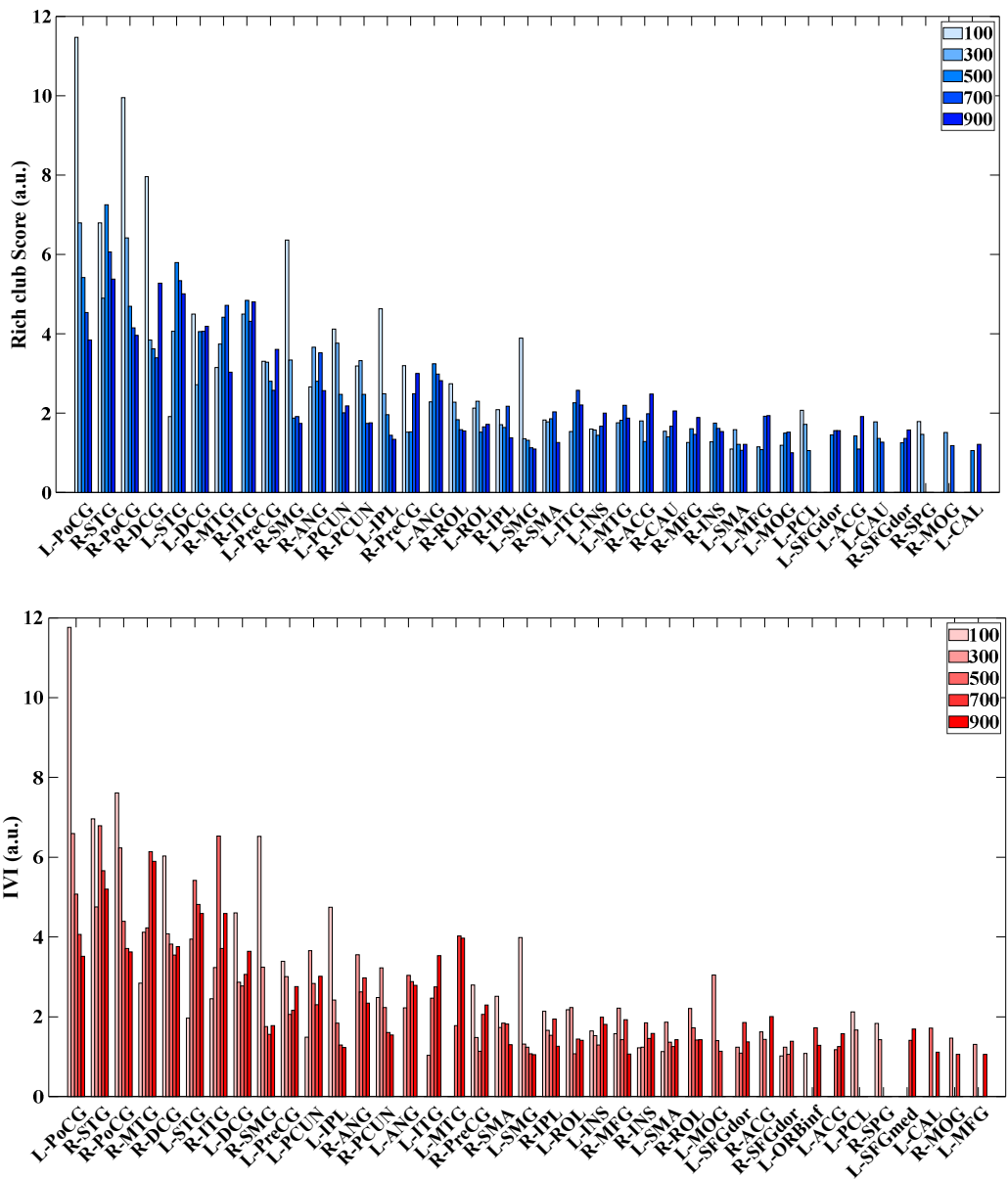

**Figure 4.** Rich-club and IVI scores of the top 20% of nodes with high RC/IVI at different nodal scales.

**Network vulnerability**

Fig. 5 shows the network vulnerability to damage to RC nodes and connections in each RSN at different scales. Among all RSNs (Fig. 5a), the functional networks showed high vulnerability to damage to RC nodes in the sensory-motor network resulting in a significant increase in network segregation (modularity and clustering coefficient) and a decrease in network integration (degree, global efficiency, and betweenness) with up to 50.7% change across all scales. The network vulnerability to loss of RC nodes in CAS, DMN and LAN was also significant with up to 19.3% increase in clustering coefficient. Lower network vulnerabilities were observed for full targeted attacks on RC nodes within SC and TMN. The highest network resilience was found for loss of RC nodes in VIS and CB mainly due to their weak involvement in the functional rich club as illustrated in Fig. 3.

The network vulnerability to attacks on connections of RC nodes was also high when the degree of RC nodes dropped below the minimum rich club level (Fig. 5b). In this scenario, RC nodes lost their rich-clubness and became local. The trend of alterations in network measures was similar to that found in the first scenario. However, the network modularity and degree were more susceptible to full targeted attacks on RC connections. Interestingly, the random attack on RC connections increased the network modularity and clustering coefficient and betweenness. To a lower extent, random attacks on local connections of local nodes, while keeping their RC and feeder connections intact, reduced the network modularity and global efficiency and increased the clustering coefficient (Fig. 5b).

Lastly, the functional brain network was relatively resilient to full targeted attacks on connections of RC nodes, while maintaining their degree above the minimum rich club (k) level (Fig. 5c). The full attack on connections of RC nodes in SM resulted in an increase (up to 14.9%) in modularity and a reduction in degree (15.2%). In all three scenarios, the network vulnerability to damage to RC connectivity was relatively higher at lower network sizes.

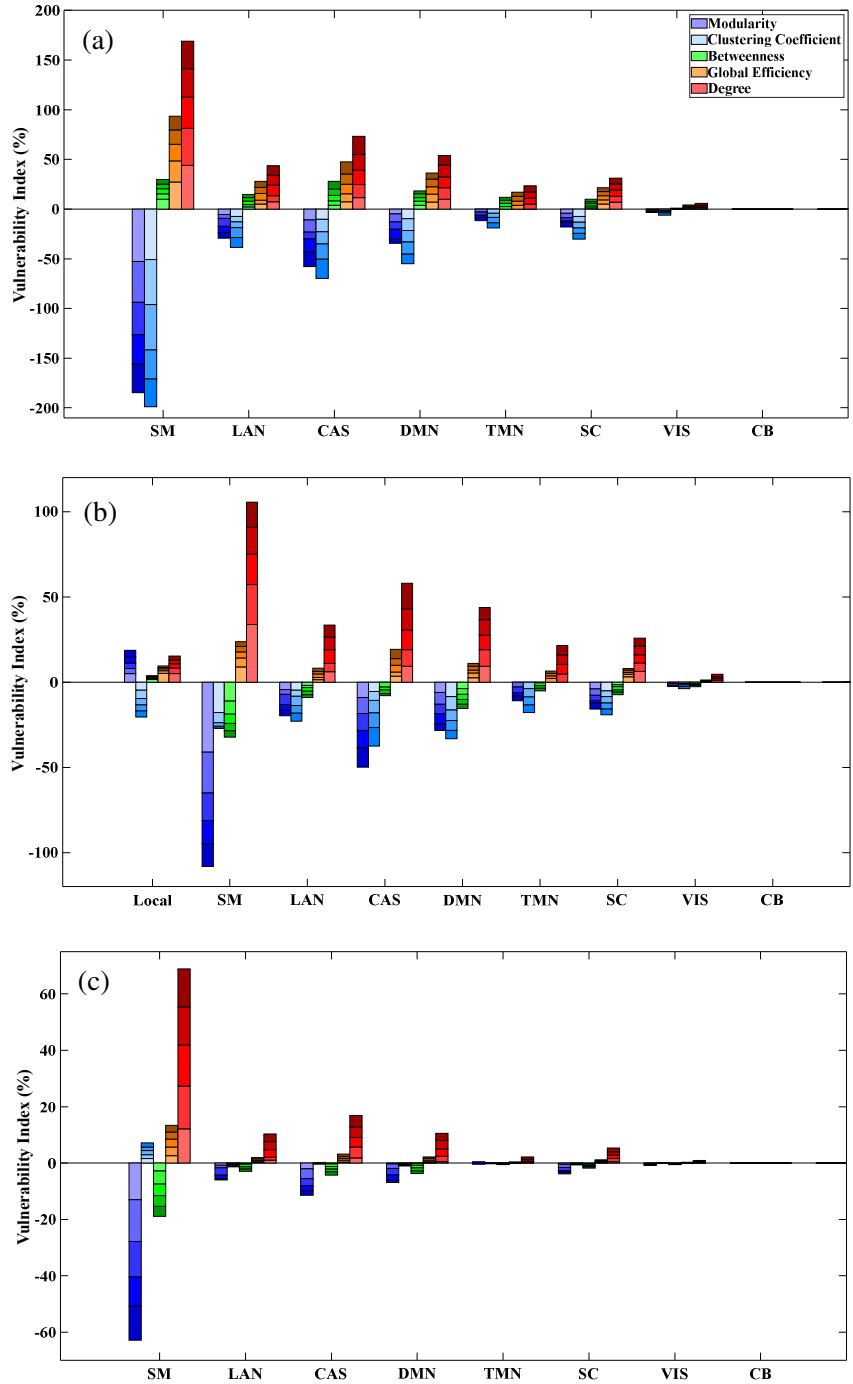

**Figure 5.** Vulnerability index computed based on different network measures for targeted and random attacks on rich club (RC) nodes and connections within different RSNs at low (lightest color) to high (darker colors) resolution nodal scales. **a.** Full targeted attacks on RC nodes, **b.** random attacks on connections of RC nodes within each RSN, where RC nodes lost their rich-clubness and became local with a degree below the minimum rich club (k) level, **c.** random attacks on connections to RC nodes, where RC nodes maintained their rich-clubness with a k level set at the lowest level within the rich club range.

**Discussion**

In this study, we performed a multiscale connectivity analysis to explore the functional organization of brain networks in full-term neonates, focusing on rich-club hubs and spreaders. We also assessed the network vulnerability to loss of rich-club nodes and connections in different functional modules. In the following sections, we discuss the general implications of our findings within the context of network properties and RC vulnerability in full-term neonates.

**Topological properties**

In line with other findings (Salvador et al. 2005; Achard et al. 2006; De Asis-Cruz et al. 2015), our results indicate a small-word functional architecture for the brain in full-term neonates in different spatial scales. The small-world topology of the brain has also been reported in younger and older infants (Fair et al. 2009; Supekar et al. 2009; Gao et al. 2011), and adults (Achard et al. 2006; Van den Heuvel et al. 2008). Our results also support findings suggesting the emergence of large-scale functional brain networks during the third trimester (Fransson et al. 2011; Scheinost et al. 2016a; Cao et al. 2017a). However, compared to structural brain networks showing an increasing trend for the small-worldness with network size in neonates and adults (Zalesky et al. 2010; Fouladivanda et al. 2021), the neonatal functional brain architecture showed significantly lower values for small-worldness, remaining relatively stable across different nodal scales. This suggests that functional brain networks at birth may be suboptimal in terms of both network segregation and integration in comparison with their respective structural infrastructure (Fouladivanda et al. 2021).

In contrast to structural connectivity patterns showing dependency on nodal scales in neonates and adults (Hagmann et al. 2008; Zalesky et al. 2010; Wei et al. 2017; Fouladivanda et al. 2021; Khalilian et al. 2021), our results showed that most of the global network metrics derived from rsfMRI data in neonates remained relatively unchanged with a network size comprising at least 300 nodes (Fornito et al. 2010; Lacy and Robinson 2020). Compared to structural brain networks in neonates (Fouladivanda et al. 2021), the brain functional architecture showed lower values for the (normalized) clustering coefficient and mean local efficiency and higher values for the mean nodal degree and betweenness centrality, which significantly increased at finer spatial resolutions. These findings suggest that the brain networks are functionally less segregated in comparison with their respective structural architecture at birth. The global and local efficiencies are shown to follow an increasing and decreasing trend with age after birth, respectively (Gozdas et al. 2019). The maturation process for functional networks is suggested to continue in terms of functional segregation and integration in the first two years of life (Gao et al. 2011). In other words, the capacity of rapid information exchange among distributed elements (Bullmore and Sporns 2012) increases with age, while the efficiency of information transfer in the immediate neighborhood of each node reduces in older ages (Latora and Marchiori 2001).

**Rich club regions**

Our results demonstrated the existence of rich-club organization in functional brain networks in full-term neonates. The rich-club organization is reported to appear at 31.3 weeks, expanding with age (Cao et al. 2017b). A high level of consistency was found across findings on functional and structural rich club organization in neonates (Grayson et al. 2014; Scheinost et al. 2016b; Fouladivanda et al. 2021). This may imply that the functional integration and segregation of information in neonates are essentially based on direct white matter pathways between rich-club regions emerging at the beginning of the third trimester, during which the number of structural connections between rich-club regions and other regions increases (Ball et al. 2014; Cao et al. 2017a).

Among all resting-state networks, our results show that the sensory-motor, cognitive-attention-salience, default mode, and language-auditory networks were largely involved in the functional rich club in neonates. Within these networks, the precentral and postcentral gyri, superior, middle and inferior temporal gyri, median cingulate, precuneus, angular gyri, anterior cingulate gyrus, middle frontal gyrus, anterior cingulate gyrus, supramarginal gyrus, supplementary motor area, rolandic operculum, inferior parietal lobule and insula were involved in the rich club at all scales. Our findings are in line with previous findings in preterm and term neonates, older infants and adults (Doria et al. 2010; Smyser et al. 2010; Fransson et al. 2011; Gao et al. 2011; van den Heuvel and Sporns 2011; Ball et al. 2014; Grayson et al. 2014; De Asis-Cruz et al. 2015; Cao et al. 2017b, a). In our study, the majority of the functional hubs also showed high spreading potential associated with their betweenness centrality, with the exception of several non-RC nodes within the cognitive-attention-salience and temporal networks that were identified as the most influential nodes (spreaders) across the whole network.

The cortical hubs in sensorimotor cortices reflect, to some extent, the role of rich club regions in perception-action at birth and provide a neuronal substrate for information processing for hubs located in higher-order association cortices typically found in adolescents and adults (Lin et al. 2008; Liu et al. 2008; Fair et al. 2009; Fransson et al. 2011; Gao et al. 2015, 2017; Dall'Orso et al. 2018).

Our findings also demonstrated the involvement of regions within high-order associative cortices and deep brain regions in the functional rich club in neonates, including the precuneus, angular gyri, middle frontal gyrus, anterior cingulate gyrus, caudate and insula (De Asis-Cruz et al. 2015). These networks are suggested to undergo rapid development in the first postnatal year, facilitating the increase of functional integration (Amsterdam 1972; Haith et al. 1988; Reznick 2009; Gao et al. 2015).

Consistent with other findings (Fransson et al. 2011), our results revealed the involvement of both the posterior and anterior parts of DMN in the functional rich club, connected through sparse connections, especially at finer spatial scales. This may support other findings suggesting the emergence of the primitive default-mode network with less mature anterior-posterior connections in neonates (Fransson et al. 2007, 2011; Gao et al. 2009; Fouladivanda et al. 2021).

In line with other studies (Ball et al. 2014), we found that rich club and feeder connections were relatively longer than those between non-rich club regions. Long-distance brain communication between functional hubs is suggested to play an important role in global brain communication required for efficient cognitive brain functioning (Van Den Heuvel et al. 2009). However, we found that the majority of functional connections were short-range local connections in neonates in comparison with structural brain networks (Fouladivanda et al. 2021), in which structural connections were dominated by long-range rich-club and feeder connections, reported to be established in the second and third trimester (Kostović and Jovanov-Milošević 2006; Fair et al. 2009; Supekar et al. 2009; Dosenbach et al. 2010; Gao et al. 2011). These findings support the idea that the brain development gradually evolves from proximity-based connections ensuring primary functions to a more distributed and integrative topology required for more complex cognitive functions (Hagmann et al. 2010; Yap et al. 2011; Bullmore and Sporns 2012; Van Den Heuvel et al. 2012; Tymofiyeva et al. 2013; Collin et al. 2014; Vértes and Bullmore 2015).

**Vulnerability analysis**

In adults, it is suggested that any damage to rich-club hubs can have a significant effect on cognition due to their central role in integrating neural information between brain regions (Van Den Heuvel et al. 2012; Collin et al. 2014). In neonates, it is shown that brain networks in neonates are more resilient to attacks on hubs than scale-free networks (De Asis-Cruz et al. 2015). In preterm babies, prematurity is reported to be associated with an immature network structure causing long-term cognitive impairments (Scheinost et al. 2016b; de Almeida et al. 2021). Our results indicated high vulnerability for the neonatal functional networks to attacks on RC nodes within the sensori-motor regions including the precentral and postcentral gyri, supramarginal gyrus, supplementary motor area, median cingulate gyri). The network vulnerability to damage to RC nodes within DMN, CAS and LAN was also significant but relatively lower than SM. Scheinost et al. (2016) have shown that the prematurity can significantly alter the functional rich club organization by reducing connections between RC nodes resulting in reduced modularity. Our results support this finding, however, when only local connections are attacked. In addition, we found that random attacks on connections of RC nodes could result in network deficiency when the degree of RC nodes dropped below the lowest rich club level. In this case, RC nodes became local and lost their integrative role. Regardless of functional modules, the functional networks showed more resilience to attacks on connections of RC nodes, when the degree of RC nodes was kept within the rich club zone at the lowest k level. Among all network segregation and integration measures, we found that the vulnerability index based on modularity and degree and, to some extent, clustering coefficient would better demonstrate the susceptibility of functional networks to attacks on RC connectivity. Overall, our findings suggest that the network integration can be highly compromised by damage to RC connectivity due to brain immaturity.

**Conclusion**

We investigated the brain functional connectome in full-term neonates using a multiscale approach, focusing on rich club hubs and spreaders. Our results revealed a rich club organization and small-world topology for functional brain networks at birth regardless of the network size, with the majority of the functional rich club nodes found to be located primarily in the sensory-motor, cognitive-attention-salience, default mode and language-auditory networks. The brain functional architecture in neonates was also found to be essentially based on short-range connections. By testing mechanistic hypotheses about disruption in RC connectivity, we found that the neonatal functional brain network was highly vulnerable to targeted and random attacks on RC hubs and connections within sensory-motor systems. The network vulnerability to loss of RC connectivity within the language and auditory, cognitive-attention-salience and default mode networks was also significant, but relatively less prominent. Our findings would help to better understand the susceptibility of the neonatal brain networks to damage caused by the brain immaturity to rich-club connectivity within and between resting-state networks.

**Acknowledgments**

"Data were provided by the developing Human Connectome Project, KCL-Imperial-Oxford Consortium funded by the European Research Council under the European Union Seventh Framework Programme (FP/2007-2013) / ERC Grant Agreement no. [319456]. The work was also supported by the NIHR Biomedical Research Centres at Guys and St Thomas' NHS Foundation Trust". This work was partially supported by the Cognitive Sciences and Technologies Council (CSTC) of Iran (Grant no. 8473) and the Ministry of Science, Research and Technology (Grant no. 18-00-02-000493).

**Competing interests**

The authors report no competing interests.

**Data Availability Statement**

The dataset analyzed in the this study is publicly available in the developing Human Connectome Project (dHCP) repository (http://www.developingconnectome.org/).

**References**

Achard S, Salvador R, Whitcher B, et al (2006) A resilient, low-frequency, small-world human brain functional network with highly connected association cortical hubs. J Neurosci 26:63–72. https://doi.org/10.1523/JNEUROSCI.3874-05.2006

Amaral LAN, Scala A, Barthelemy M, Stanley HE (2000) Classes of small-world networks. Proc Natl Acad Sci 97:11149–11152. https://doi.org/https://doi.org/10.1073/pnas.200327197

Amsterdam B (1972) Mirror self-image reactions before age two. Dev Psychobiol J Int Soc Dev Psychobiol 5:297–305

Ball G, Aljabar P, Zebari S, et al (2014) Rich-club organization of the newborn human brain. Proc Natl Acad Sci U S A 111:7456–7461. https://doi.org/https://doi.org/10.1073/pnas.1324118111

Bassett DS, Bullmore ED (2006) Small-world brain networks. Neurosci 12:512–523

Bassett DS, Bullmore ET, Meyer-Lindenberg A, et al (2009) Cognitive fitness of cost-efficient brain functional networks. Proc Natl Acad Sci 106:11747–11752. https://doi.org/https://doi.org/10.1073/pnas.0903641106

Bruchhage MMK, Ngo G-C, Schneider N, et al (2020) Functional connectivity correlates of infant and early childhood cognitive development. Brain Struct Funct 225:669–681

Buckner RL, Krienen FM (2013) The evolution of distributed association networks in the human brain. Trends Cogn

Sci 17:648–665

Bullmore E, Sporns O (2009) Complex brain networks: Graph theoretical analysis of structural and functional systems (Nature Reviews Neuroscience (2009) 10, (186-198)). Nat. Rev. Neurosci. 10:312

Bullmore E, Sporns O (2012) The economy of brain network organization. Nat Rev Neurosci 13:336–349. https://doi.org/https://doi.org/10.1038/nrn3214

Calhoun VD, Adali T, Pearlson GD, Pekar JJ (2001) A method for making group inferences from functional MRI data using independent component analysis. Hum Brain Mapp 14:140–151. https://doi.org/10.1002/hbm.1048

Cao M, He Y, Dai Z, et al (2017a) Early development of functional network segregation revealed by connectomic analysis of the preterm human brain. Cereb Cortex 27:1949–1963. https://doi.org/10.1093/cercor/bhw038

Cao M, Huang H, He Y (2017b) Developmental Connectomics from Infancy through Early Childhood. Trends Neurosci. 40:494–506

Chen D-B, Gao H, Lü L, Zhou T (2013) Identifying influential nodes in large-scale directed networks: the role of clustering. PLoS One 8:e77455

Chklovskii DB, Schikorski T, Stevens CF (2002) Wiring optimization in cortical circuits. Neuron 34:341–347

Collin G, Sporns O, Mandl RCW, Van Den Heuvel MP (2014) Structural and functional aspects relating to cost and benefit of rich club organization in the human cerebral cortex. Cereb Cortex 24:2258–2267. https://doi.org/https://doi.org/10.1093/cercor/bht064

Costa L da F, Rodrigues FA, Travieso G, Villas Boas PR (2007) Characterization of complex networks: A survey of measurements. Adv Phys 56:167–242

Craddock RC, James GA, Holtzheimer PE, et al (2012) A whole brain fMRI atlas generated via spatially constrained spectral clustering. Hum Brain Mapp 33:1914–1928. https://doi.org/10.1002/hbm.21333

Dall'Orso S, Steinweg J, Allievi AG, et al (2018) Somatotopic mapping of the developing sensorimotor cortex in the preterm human brain. Cereb cortex 28:2507–2515

de Almeida JS, Meskaldji D-E, Loukas S, et al (2021) Preterm birth leads to impaired rich-club organization and fronto-paralimbic/limbic structural connectivity in newborns. Neuroimage 225:117440

De Asis-Cruz J, Bouyssi-Kobar M, Evangelou I, et al (2015) Functional properties of resting state networks in healthy full-term newborns. Sci Rep 5:1–15. https://doi.org/10.1038/srep17755

Doria V, Beckmann CF, Arichi T, et al (2010) Emergence of resting state networks in the preterm human brain. Proc Natl Acad Sci 107:20015–20020. https://doi.org/https://doi.org/10.1073/pnas.1007921107

Dosenbach NUF, Nardos B, Cohen AL, et al (2010) Prediction of individual brain maturity using fMRI. Science (80- ) 329:1358–1361. https://doi.org/10.1126/science.1194144

Drakesmith M, Caeyenberghs K, Dutt A, et al (2015) Overcoming the effects of false positives and threshold bias in graph theoretical analyses of neuroimaging data. Neuroimage 118:313–333

Erhardt EB, Rachakonda S, Bedrick EJ, et al (2011) Comparison of multi-subject ICA methods for analysis of fMRI data. Hum Brain Mapp 32:2075–2095. https://doi.org/10.1002/hbm.21170

Fair DA, Cohen AL, Power JD, et al (2009) Functional brain networks develop from a "local to distributed" organization. PLoS Comput biol 5:e1000381. https://doi.org/10.1371/journal.pcbi.1000381

Fitzgibbon SP, Harrison SJ, Jenkinson M, et al (2020) The developing Human Connectome Project (dHCP) automated resting-state functional processing framework for newborn infants. Neuroimage 223:117303. https://doi.org/https://doi.org/10.1016/j.neuroimage.2020.117303

Fornito A, Zalesky A, Bullmore ET (2010) Network scaling effects in graph analytic studies of human resting-state FMRI data. Front Syst Neurosci 4:22

Fouladivanda M, Kazemi K, Makki M, et al (2021) Multi-scale structural rich-club organization of the brain in full-term newborns: a combined DWI and fMRI study. J Neural Eng 18:46065

Fransson P, Åden U, Blennow M, Lagercrantz H (2011) The functional architecture of the infant brain as revealed by resting-state fMRI. Cereb Cortex 21:145–154. https://doi.org/10.1093/cercor/bhq071

Fransson P, Skiöld B, Horsch S, et al (2007) Resting-state networks in the infant brain. Proc Natl Acad Sci U S A 104:15531–15536. https://doi.org/10.1073/pnas.0704380104

Freeman LC, Borgatti SP, White DR (1991) Centrality in valued graphs: A measure of betweenness based on network flow. Soc Networks 13:141–154

Freeman LC, Roeder D, Mulholland RR (1979) Centrality in social networks: II. Experimental results. Soc Networks 2:119–141

Fukushima M, Betzel RF, He Y, et al (2018) Structure–function relationships during segregated and integrated network states of human brain functional connectivity. Brain Struct Funct 223:1091–1106

Gao W, Alcauter S, Smith JK, et al (2015) Development of human brain cortical network architecture during infancy. Brain Struct Funct 220:1173–1186

Gao W, Gilmore JH, Giovanello KS, et al (2011) Temporal and spatial evolution of brain network topology during the first two years of life. PLoS One 6:e25278

Gao W, Lin W, Grewen K, Gilmore JH (2017) Functional connectivity of the infant human brain: plastic and modifiable. Neurosci 23:169–184. https://doi.org/https://doi.org/10.1177/1073858416635986

Gao W, Zhu H, Giovanello KS, et al (2009) Evidence on the emergence of the brain's default network from 2-week-old to 2-year-old healthy pediatric subjects. Proc Natl Acad Sci 106:6790–6795

Goelman G, Gordon N, Bonne O (2014) Maximizing negative correlations in resting-state functional connectivity MRI by time-lag. PLoS One 9:e111554

Gozdas E, Holland SK, Altaye M (2019) Developmental changes in functional brain networks from birth through adolescence. Hum Brain Mapp 40:1434–1444. https://doi.org/10.1002/hbm.24457

Grayson DS, Ray S, Carpenter S, et al (2014) Structural and functional rich club organization of the brain in children and adults. PLoS One 9:e88297. https://doi.org/10.1371/journal.pone.0088297

Guillery RW (2005) Is postnatal neocortical maturation hierarchical? Trends Neurosci 28:512–517

Hagmann P, Cammoun L, Gigandet X, et al (2008) Mapping the structural core of human cerebral cortex. PLoS Biol 6:1479–1493. https://doi.org/10.1371/journal.pbio.0060159

Hagmann P, Sporns O, Madan N, et al (2010) White matter maturation reshapes structural connectivity in the late developing human brain. Proc Natl Acad Sci U S A 107:19067–19072. https://doi.org/https://doi.org/10.1073/pnas.1009073107

Haith MM, Hazan C, Goodman GS (1988) Expectation and anticipation of dynamic visual events by 3.5-month-old babies. Child Dev 467–479

Howell AL, Osher DE, Li J, Saygin ZM (2020) The intrinsic neonatal hippocampal network: rsfMRI findings. J Neurophysiol 124:1458–1468. https://doi.org/https://doi.org/10.1152/jn.00362.2020

Hughes EJ, Winchman T, Padormo F, et al (2017) A dedicated neonatal brain imaging system. Magn Reson Med 78:794–804. https://doi.org/https://doi.org/10.1002/mrm.26462

Kaiser M, Hilgetag CC (2006) Nonoptimal component placement, but short processing paths, due to long-distance projections in neural systems. PLoS Comput Biol 2:e95

Khalilian M, Kazemi K, Fouladivanda M, et al (2021) Effect of Multishell Diffusion MRI Acquisition Strategy and Parcellation Scale on Rich-Club Organization of Human Brain Structural Networks. Diagnostics 11:970

Kitsak M, Gallos LK, Havlin S, et al (2010) Identification of influential spreaders in complex networks. Nat Phys 6:888–893

Kostović I, Jovanov-Milošević N (2006) The development of cerebral connections during the first 20-45 weeks' gestation. Semin Fetal Neonatal Med 11:415–422. https://doi.org/10.1016/j.siny.2006.07.001

Lacy TC, Robinson PA (2020) Effects of parcellation and threshold on brainconnectivity measures. PLoS One 15:e0239717

Latora V, Marchiori M (2001) Efficient behavior of small-world networks. Phys Rev Lett 87:198701. https://doi.org/10.1103/PhysRevLett.87.198701

Lin W, Zhu Q, Gao W, et al (2008) Functional connectivity MR imaging reveals cortical functional connectivity in the developing brain. Am J Neuroradiol 29:1883–1889. https://doi.org/10.3174/ajnr.A1256

Liu W-C, Flax JF, Guise KG, et al (2008) Functional connectivity of the sensorimotor area in naturally sleeping infants. Brain Res 1223:42–49

Lü L, Zhou T, Zhang Q-M, Stanley HE (2016) The H-index of a network node and its relation to degree and coreness. Nat Commun 7:10168. https://doi.org/10.1038/ncomms10168

Makropoulos A, Robinson EC, Schuh A, et al (2018) The developing human connectome project: A minimal processing pipeline for neonatal cortical surface reconstruction. Neuroimage 173:88–112. https://doi.org/https://doi.org/10.1016/j.neuroimage.2018.01.054

Maslov S, Sneppen K (2002) Specificity and stability in topology of protein networks. Science (80- ) 296:910–913. https://doi.org/10.1126/science.1065103

Morone F, Makse HA (2015) Influence maximization in complex networks through optimal percolation. Nature 524:65–68

Newman MEJ (2002) Assortative mixing in networks. Phys Rev Lett 89:208701

Newman MEJ (2006) Modularity and community structure in networks. Proc Natl Acad Sci 103:8577–8582

Price A, Cordero-Grande L, Malik S, et al (2015) Accelerated neonatal fMRI using multiband EPI. In: Proceedings of the 23rd Annual Meeting of ISMRM, Toronto, Canada. p 3911

Rachakonda S, Egolf E, Correa N, Calhoun V (2007) Group ICA of fMRI toolbox (GIFT) manual

Reznick JS (2009) Working memory in infants and toddlers.

Salavaty A, Ramialison M, Currie PD (2020) Integrated value of influence: an integrative method for the identification of the most influential nodes within networks. Patterns 1:100052

Salimi-Khorshidi G, Douaud G, Beckmann CF, et al (2014) Automatic denoising of functional MRI data: combining independent component analysis and hierarchical fusion of classifiers. Neuroimage 90:449–468. https://doi.org/10.1016/j.neuroimage.2013.11.046

Salvador R, Suckling J, Coleman MR, et al (2005) Neurophysiological architecture of functional magnetic resonance images of human brain. Cereb Cortex 15:1332–2342. https://doi.org/10.1093/cercor/bhi016

Scheinost D, Kwon SH, Lacadie C, et al (2016a) Prenatal stress alters amygdala functional connectivity in preterm neonates. NeuroImage Clin 12:381–388. https://doi.org/10.1016/j.nicl.2016.08.010

Scheinost D, Kwon SH, Shen X, et al (2016b) Preterm birth alters neonatal, functional rich club organization. Brain Struct Funct. https://doi.org/10.1007/s00429-015-1096-6

Serag A, Aljabar P, Ball G, et al (2012) Construction of a consistent high-definition spatio-temporal atlas of the developing brain using adaptive kernel regression. Neuroimage 59:2255–2265. https://doi.org/https://doi.org/10.1016/j.neuroimage.2011.09.062

Smith SM, Fox PT, Miller KL, et al (2009) Correspondence of the brain's functional architecture during activation and rest. Proc Natl Acad Sci 106:13040–13045

Smyser CD, Inder TE, Shimony JS, et al (2010) Longitudinal analysis of neural network development in preterm infants. Cereb Cortex 20:2852–2862. https://doi.org/10.1093/cercor/bhq035

Sporns O (2010) Networks of the Brain. MIT press

Sporns O, Tononi G, Kötter R (2005) The human connectome: A structural description of the human brain. PLoS Comput. Biol. 1:0245–0251

Supekar K, Musen M, Menon V (2009) Development of large-scale functional brain networks in children. PLoS Biol 7:e1000157. https://doi.org/10.1371/journal.pbio.1000157

Thomas Yeo BT, Krienen FM, Sepulcre J, et al (2011) The organization of the human cerebral cortex estimated by intrinsic functional connectivity. J Neurophysiol 106:1125–1165. https://doi.org/10.1152/jn.00338.2011

Thompson WH, Fransson P (2015) The frequency dimension of fMRI dynamic connectivity: Network connectivity, functional hubs and integration in the resting brain. Neuroimage 121:227–242

Tymofiyeva O, Hess CP, Ziv E, et al (2013) A DTI-Based Template-Free Cortical Connectome Study of Brain Maturation. PLoS One 8:1–10. https://doi.org/10.1371/journal.pone.0063310

Tzourio-Mazoyer N, Landeau B, Papathanassiou D, et al (2002) Automated anatomical labeling of activations in SPM using a macroscopic anatomical parcellation of the MNI MRI single-subject brain. Neuroimage 15:273–289. https://doi.org/10.1006/nimg.2001.0978

Van Den Heuvel MP, Kahn RS, Goñi J, Sporns O (2012) High-cost, high-capacity backbone for global brain communication. Proc Natl Acad Sci U S A 109:11372–11377. https://doi.org/10.1073/pnas.1203593109

Van Den Heuvel MP, Kersbergen KJ, De Reus MA, et al (2015) The neonatal connectome during preterm brain development. Cereb Cortex 25:3000–3013. https://doi.org/10.1093/cercor/bhu095

Van Den Heuvel MP, Pol HEH (2010) Exploring the brain network: a review on resting-state fMRI functional connectivity. Eur Neuropsychopharmacol 20:519–534

van den Heuvel MP, Sporns O (2013) An anatomical substrate for integration among functional networks in human cortex. J Neurosci 33:14489–14500. https://doi.org/10.1523/JNEUROSCI.2128-13.2013

van den Heuvel MP, Sporns O (2011) Rich-club organization of the human connectome. J Neurosci 31:15775–15786. https://doi.org/10.1523/JNEUROSCI.3539-11.2011

Van den Heuvel MP, Stam CJ, Boersma M, Pol HEH (2008) Small-world and scale-free organization of voxel-based resting-state functional connectivity in the human brain. Neuroimage 43:528–539. https://doi.org/10.1016/j.neuroimage.2008.08.010

Van Den Heuvel MP, Stam CJ, Kahn RS, Pol HEH (2009) Efficiency of functional brain networks and intellectual performance. J Neurosci 29:7619–7624. https://doi.org/https://doi.org/10.1523/JNEUROSCI.1443-09.2009

Vértes PE, Bullmore ET (2015) Annual research review: growth connectomics–the organization and reorganization of brain networks during normal and abnormal development. J Child Psychol Psychiatry 56:299–320

Watts DJ, Strogatz SH (1998) Collective dynamics of 'small-world9 networks. Nature 393:440–442. https://doi.org/https://doi.org/10.1038/30918

Wei K, Cieslak M, Greene C, et al (2017) Sensitivity analysis of human brain structural network construction. Netw Neurosci 1:446–467. https://doi.org/10.1162/NETN_a_00025

Yap P-T, Fan Y, Chen Y, et al (2011) Development trends of white matter connectivity in the first years of life. PLoS One 6:e24678. https://doi.org/10.1371/journal.pone.0024678

Zalesky A, Fornito A, Harding IH, et al (2010) Whole-brain anatomical networks: Does the choice of nodes matter? Neuroimage 50:970–983. https://doi.org/10.1016/j.neuroimage.2009.12.027